\documentclass[journal,twoside]{IEEEtran}
\ifCLASSINFOpdf
\else
\fi
\usepackage{amsmath,amssymb,amsfonts}
\usepackage{algorithmic}
\usepackage{graphicx}
\usepackage{textcomp}

\usepackage{hyperref}
\usepackage{amsfonts, amsmath, amssymb}
\usepackage{graphicx}
\usepackage{afterpage}
\usepackage{multirow}
\usepackage{enumerate}
\usepackage{subcaption}
\usepackage{epstopdf}

\DeclareMathOperator{\R}{{\mathbb R}}
\DeclareMathOperator{\C}{{\mathbb C}}

\DeclareMathOperator{\bndry}{\partial\Omega}
\DeclareMathOperator{\texp}{\mathbf{t}^{\mbox{\tiny{\textbf{exp}}}}}
\DeclareMathOperator{\texpND}{\mathbf{t}^{\mbox{\tiny{\textbf{exp}}}}_{\mbox{\tiny{ND}}}}
\DeclareMathOperator{\texpNDdiff}{\mathbf{t}^{\mbox{\tiny{\textbf{exp,diff}}}}_{\mbox{\tiny{ND}}}}
\DeclareMathOperator{\sigdiff}{\sigma^{\mbox{\tiny{\textbf{diff}}}}}

\DeclareMathOperator{\qexp}{\mathbf{q}^{\mbox{\tiny{\textbf{exp}}}}}
\DeclareMathOperator{\sigcal}{\sigma^{\mbox{\tiny{\textbf{CAL}}}}}

\DeclareMathOperator{\sigexp}{\sigma^{\mbox{\tiny{\textbf{exp}}}}}

\DeclareMathOperator{\Fhat}{\hat{F}}
\DeclareMathOperator{\FhatND}{\hat{F}_{\mbox{\tiny{ND}}}}
\DeclareMathOperator{\FhatNDdiff}{\hat{F}^{\mbox{\tiny{\textbf{diff}}}}_{\mbox{\tiny{ND}}}}

\DeclareMathOperator{\sigBest}{\sigma_{\mbox{\tiny{best}}}}

\begin{document}

%
\title{Fast 3D Partial Boundary Data EIT Reconstructions using Direct Inversion CGO-based Methods}
%
%
%
\author{S.~J. Hamilton, P.~Muller, V.~Kolehmainen, J.~Toivanen
\thanks{Manuscript submitted: December 13, 2024}
\thanks{SH was supported by the National Institute Of Biomedical Imaging and Bioengineering of the National Institutes of Health under Award Number R21EB028064. The content is solely the responsibility of the authors and does not necessarily represent the official views of the National Institutes of Health. JT and VK were supported by the Research council of Finland (Projects No. 359433, 353084, Finnish Centre of Excellence in Inverse Modeling and Imaging; and Projects No. 358944, Flagship of Advanced Mathematics for Sensing Imaging and Modelling; and the Jane and Aatos Erkko Foundation.}
\thanks{S.J. Hamilton is with the Department of Mathematical and Statistical Sciences; Marquette University, Milwaukee, WI 53233 USA, \texttt{(sarah.hamilton@mu.edu)}}%
\thanks{P.~A. Muller is with the Department of Mathematics \& Statistics; Villanova University, Villanova, PA 19085 USA,  email: \texttt{peter.muller@villanova.edu}}
\thanks{V. Kolehmainen and J. Toivanen are with the Department of Technical Physics, University of Eastern Finland, FI-70210 Kuopio, Finland, email: \texttt{ville.kolehmainen@uef.fi, jussi.toivanen@uef.fi}}}%

\maketitle

\begin{abstract}
The first partial boundary data complex geometrical optics based methods for electrical impedance tomography in three dimensions are developed, and tested, on simulated and experimental data.  The methods provide good localization of targets for both absolute and time-difference imaging, when large portions of the domain are inaccessible for measurement.  As most medical applications of electrical impedance tomography are limited to partial boundary data, the development of partial boundary algorithms is highly desirable.  While iterative schemes have been used traditionally, their high computational cost makes them cost-prohibitive for applications that need fast imaging. The proposed algorithms require no iteration and provide informative absolute or time-difference images exceptionally quickly in under 2 seconds.  Reconstructions are compared to reference reconstructions from standard linear difference imaging (30 seconds) and total variation regularized absolute imaging (several minutes)  The algorithms perform well under high levels of noise and incorrect domain modeling. 
\end{abstract}



%
\IEEEpeerreviewmaketitle

\section{Introduction}\label{sec:Intro}
\IEEEPARstart{E}{lectrical} impedance tomography (EIT) is a relatively new, portable, inexpensive, radiation-free imaging modality with potential for a range of medical applications, including on-line monitoring of lung function and post-stroke follow-in intensive care units \cite{deCastroMartins2019,toivanen2024simulation}.  
In these applications, fast imaging and image reconstruction is essential as the monitoring data can be used, for example, to guide mechanical ventilation or urgent treatment decisions. Time difference imaging is the most effective way to monitor changes in these applications, and complex geometrical optics (CGO) direct inversion techniques have great promise for \emph{fast} imaging, as they can produce 3D reconstruction in just few seconds. CGO-based methods also have also been shown to have good robustness to noise and domain modeling errors \cite{Hamilton2021}. Absolute images are beneficial for detection, such as of stroke or cancer, for which reference data may not be available, and CGO-based methods are equally fast for absolute imaging, whereas traditional methods are not.

The continuum model for EIT follows the conductivity equation within a domain $\Omega\subset\mathbb{R}^n$ with $n\geq2$, as derived from quasi-static assumptions on Maxwell's equations
\begin{align}
\begin{split}\label{eq:cond}
\nabla\cdot \sigma(x)\nabla u(x)  =  0,& \quad x\in\Omega\\
\left.\sigma\frac{\partial u}{\partial\nu}\right|_{\partial\Omega} = f(x), \,\,u(x)|_{\partial\Omega}=g(x),& \quad x\in\bndry,
\end{split}
\end{align}
where $u(x)$ is the electric potential, $\sigma(x)\in L^\infty(\Omega)$ the isotropic conductivity, $f(x)$ the applied current density on the boundary, and $g(x)$ the resulting measured voltage on the boundary. The conductivity $\sigma(x)$ is satisfies $\frac1C\leq \sigma\leq C$ for some constant $C$. These boundary conditions can be described by the Neumann-to-Dirichlet (ND) map $\mathcal{R}_\sigma: f(x) \mapsto g(x)$. The goal of EIT is to recover $\sigma(x)$ from the ND map or its inverse, the Dirichlet-to-Neumann (DN) map, $\Lambda_\sigma: g(x)\mapsto f(x)$, if voltages are applied and currents are measured. In practice, these infinite precision maps are approximated by applying several linearly independent current patterns on finitely many electrodes around the domain and measuring the resulting voltages from each applied current pattern. 

In medical imaging, the inverse conductivity problem of EIT is inherently a three-dimensional problem as even with planar arrays of electrodes around a domain, applied currents are known travel outside of those planar slices.  Hence, it is natural to pose the recovery of the conductivity in a 3D setting.  For iterative/optimization-based reconstruction schemes the task of determining the interior conductivity that would lead to the best-fit of the measured voltage data for given applied currents, subject to a desirable regularization term, is unchanged.  The absence, or presence, of full boundary data coverage is irrelevant.  While one cannot expect to accurately recover the interior conductivity far from the electrodes, nearby such approaches perform adequately.  However, in the (CGO)-based reconstruction setting \cite{Nachman1988a, Delbary2014, Hamilton2021, Hamilton2022}, the main first step of the approach, using the measured current/voltage data to compute the non-physical scattering data, involves an integral over the boundary of the domain of interest.  The same is true of the linearized CGO-based Calder\'on's method \cite{Calderon1980, Hamilton2021, Hamilton2022}, where the Fourier transform of the image is recovered from an integral over the boundary of the domain of interest. This boundary integral results, in part, from integration by parts, moving the problem from an interior problem to a surface problem.  Hence the integral around the domain cannot be easily avoided with these CGO-based frameworks.  

A further challenge is that the theoretical foundation of the 3D CGO-based methods \cite{Nachman1988a, Novikov1987} was developed for the DN map.  As an approximation to the ND map is more commonly measured experimentally, the matrix approximation to the ND map is inverted to achieve the matrix approximation for the DN map.  With partial boundary data, this becomes impractical.  To bypass this challenge, building upon the partial data approach for 2D D-bar (CGO) methods for EIT, \cite{Hauptmann2017, Hauptmann2017a}, we develop an ND based framework for the CGO-based reconstruction methods in 3D.

Regarding novelty, this work is the first to derive, and use, 3D CGO-based methods formulated using the ND map for either full or partial boundary data. It provides the first formulation of Calder\'on's method in the ND framework in any dimension and the first to reconstruct absolute images in 3D from partial boundary data using CGO-based methods. Calder\'on's method has previously been used to reconstruct 3D difference images from partial boundary data using the DN framework in cylindrical and torso-like domains \cite{Shin2020,Shin2024}, but the $\texp$ method used here has never been used for partial boundary data.

The remainder of the paper is organized as follows.  Section~\ref{sec:Methods} presents the new CGO-based reconstructions methods for partial boundary data and reviews classic methods that will be used for comparison.  Section~\ref{sec:metricstesting} describes the simulated and experimental data, as well as the evaluation metrics that will be used to assess the quality of the reconstructions.  Results are presented and discussed in Section~\ref{sec:results} and conclusions drawn in Section~\ref{sec:conclusion}.

\section{Methods}\label{sec:Methods}
In this work we develop two novel CGO-based reconstruction methods for the partial boundary data problem by formulating them directly in the ND, rather than DN, setting. 
\subsection{The $\texp$ Method}\label{sec:texp}
The theory behind the $\texp$ method is based on inverse scattering theory, complex geometrical optics, and D-bar methods for the associated Schr\"odinger equation (see e.g., \cite{Beals1985}). Following \cite{Nachman1988}, we work with $\sigma\in C^{1,1}(\bar{\Omega})$ with positive lower bound and where $\Omega$ is a bounded domain with a $C^{1,1}$ boundary and $\R^3\setminus \bar{\Omega}$ connected.  Applying the change of variables $\tilde{u}(x) = u(x)\sqrt{\sigma(x)}$ and $q(x) = \frac{\Delta \sqrt{\sigma(x)}}{\sqrt{\sigma(x)}}$, to the conductivity equation \eqref{eq:cond} results in
\[\nabla\cdot \sigma(x)\nabla u(x)  =  0 \longrightarrow [-\Delta + q(x)] \tilde{u}(x) = 0.\]

By introducing a nonphysical auxiliary parameter $\zeta\in\C^3$, Nachman \cite{Nachman1988a} and Novikov \cite{Novikov1987} independently showed that special solutions, called {\it complex geometrical optics solutions}, exist to 
\begin{equation}\label{eq:schro}
    [-\Delta + q(x)] \Psi(x,\zeta) = 0,
\end{equation}
where $\Psi(x,\zeta)\sim e^{ix\cdot\zeta}$ for large $|x|$ or $|\zeta|$, and $\zeta$ 
belongs to a subspace $V_\xi$ of $\C^3$ defined by special orthogonality properties 
{\small
\begin{equation}\label{eq:Vxi}
\mathcal{V}_\xi = \left\{\zeta\in\C^3\middle|\zeta\cdot\zeta=0, \left(\xi+\zeta\right)\cdot \left(\xi+\zeta\right)=0, \xi\in\R^3 \right\}.
\end{equation}
}
The associated \emph{scattering transform} is defined by
\begin{equation}\label{eq:t-scat-full}
t(\xi,\zeta)=\int_{\R^3} e^{-ix\cdot(\xi+\zeta)}q(x)\psi(x,\zeta)\;dx.
\end{equation}
When $|\zeta|$ is large enough, the scattering transform is  the Fourier transform of the potential $q(x)$, which allows the conductivity to be recovered via an inverse Fourier transform 
{\small
\begin{equation}\label{eq:t-q}
q(x)= \mathcal{F}^{-1}\left\{\widehat{q}(\xi)\right\}\approx \frac{1}{(2\pi)^3}\int_{\R^3} e^{i x\cdot\xi} t(\xi,\zeta)\;d\xi, \quad x\in\R^3.
\end{equation}
}
The conductivity can then be recovered from the Schr\"odinger potential $q(x)$ by solving the boundary value problem 
\begin{equation}\label{eq:t-QtoSigma}
\left\{\begin{array}{rclcl}
(-\Delta + q(x)) \tilde{u}(x) & =& 0 & \qquad & x\in\Omega\subset\R^3\\
\tilde{u}(x) & =& 1 && x\in\bndry
\end{array}
\right.
\end{equation}
and computing $\sigma(x)=\left(\tilde{u}(x)\right)^2$. However, this formulation of the scattering data \eqref{eq:t-scat-full} requires impractical knowledge of $q(x)$, and thus $\sigma(x)$, for all $x\in\R^3$.  Integration by parts connects \eqref{eq:t-scat-full} back to the measured voltage and current data, in the form of the DN map $\Lambda_\sigma$, 
\begin{equation}\label{eq:t-scat-tBIE-DN}
t(\xi,\zeta)=\int_{\bndry} e^{-ix\cdot(\xi+\zeta)}\left(\Lambda_\gamma-\Lambda_1\right)\psi(x,\zeta)\;dS(x),
\end{equation}
for $\xi\in \R^3$ and $\zeta\in V_\xi$, where the CGO traces $\psi$ on $\bndry$ can be determined from the DN map via the weakly singular Fredholm integral equation of the second kind
{\small
\begin{equation}\label{eq:psi_boundaryDN}
    \psi(x,\zeta) = e^{i x\cdot\zeta} - \int_{\bndry}G_\zeta(x-y)\left(\Lambda_\sigma-\Lambda_1 \right)\psi(y,\zeta)\;dS(y),  
\end{equation}
}
for $x\in\bndry$ and $\zeta\in V_\xi$, where $G_\zeta(x)$ denotes the Faddeev Green's function \cite{Faddeev1966}.  Note that $\Lambda_1$ is the DN map for a constant conductivity of $1$ on the same domain $\Omega$, and is either computed analytically, or simulated (e.g. by solving a Finite Element problem) with $\sigma\equiv 1$. Therefore, the conductivity $\sigma$ can be uniquely recovered from infinite precision knowledge of the DN map $\Lambda_\sigma$ via the following algorithm:
\[
\left(\Lambda_\sigma,\Lambda_1\right) \overset{(i)}{\longrightarrow}\left.\psi(x,\zeta)\right|_{\bndry} 
 \overset{(ii)}{\longrightarrow} t(\xi,\zeta)  \overset{(iii)}{\longrightarrow} q(x)  \overset{(iv)}{\longrightarrow} \sigma(x).
\]

In practice, current is applied and the resulting voltage measured.  These measurements correspond to an approximation to the ND, rather than DN, map.  When working with full boundary data, the matrix approximation to the ND map is numerically inverted to form an approximation to the DN map.  However, when working with partial boundary data, for example brain or breast imaging applications, there can be large sections of the domain about which there is no measurement data.  Therefore, one cannot expect the inversion of the ND matrix to yield the true corresponding partial boundary data DN map.  Therefore, we instead derive a version of the scattering data \eqref{eq:t-scat-tBIE-DN} directly in terms of the ND map $\mathcal{R}_\sigma$.  Once the scattering data is in hand, the remainder of the algorithm can remain the same as the earlier DN version.  

The approach is similar to that of the ND formulation in 2D~\cite{Hauptmann2017}.  Consider the following two Schr\"odinger equations with Neumann boundary conditions
\begin{equation}\label{eq:schrodinger}
    \begin{array}{rcll}
    \left(-\Delta + q_j\right)w_j &=& 0  & \in \Omega, \qquad j=1,2.\\
    \partial_\nu w_j &=&\varphi_j & \text{on} \bndry,
    \end{array}
\end{equation}
where $\nu$ is the outward facing normal to the boundary.  Alessandrini's identity \cite{Alessandrini1988} states
\begin{equation}\label{eq:aless}
    \int_\Omega \left(q_1-q_2\right) w_1 w_2 \;dz = -\int_{\bndry} \varphi_1\left(\mathcal{R}_{q_1} - \mathcal{R}_{q_2}\right)\varphi_2 dS.
\end{equation}
Let $q_1=0$ and $q_2=q$ be two potentials satisfying \eqref{eq:schrodinger}, then the harmonic function $w_1=e^{-ix\cdot(\xi+\zeta)}$ is a solution with $q_1=0$.  Let $w_2=\Psi$ be the solution corresponding to $q_2=q$.  Then
{\footnotesize
\begin{equation}\label{eq:Aless_scat}
\underset{\Omega}{\int} qe^{-ix\cdot(\xi+\zeta)} \Psi dz = \underset{\bndry}{\int} \partial_\nu \left( e^{-ix\cdot(\xi+\zeta)}\right) \left(\mathcal{R}_0 - \mathcal{R}_q\right)\partial_\nu \left(\Psi\right)dS.
\end{equation}
}\normalsize
{\noindent}Next, recognizing the left hand side of \eqref{eq:Aless_scat} as \eqref{eq:t-scat-full}, we see
\begin{equation}\label{eq:scatBIE_ND}
    t(\xi,\zeta)= \int_{\bndry} \partial_\nu \left( e^{-ix\cdot(\xi+\zeta)}\right) \left(\mathcal{R}_0 - \mathcal{R}_q\right)\partial_\nu \left(\Psi\right)dS,
\end{equation}
giving a connection from the scattering data to the measured voltage/current data this time in terms of the ND map.  Using a `Born Approximation' to the scattering data, by replacing the CGO solutions $\psi$ with their asymptotic behavior $e^{ix\cdot\xi}$ gives

{\footnotesize 
\begin{IEEEeqnarray}{rCl}
    \texpND(\xi,\zeta) &=& \int_{\bndry} \partial_\nu \left( e^{-ix\cdot(\xi+\zeta)}\right) \left(\mathcal{R}_0 - \mathcal{R}_q\right)\partial_\nu \left( e^{ix\cdot\xi} \right)dS\nonumber\\
&=& \underset{\bndry}{\int} e^{-ix\cdot(\xi+\zeta)} (\xi+\zeta)\cdot\nu \left(\mathcal{R}_q - \mathcal{R}_0\right) e^{ix\cdot\xi}(\zeta\cdot\nu)dS.\label{eq:texp_ND_q}    
\end{IEEEeqnarray}}
To relate the problem back from the Schr\"odinger potential $q$ to the conductivity $\sigma$  we use the assumption that the conductivity $\sigma=1$ in a neighborhood of $\bndry$, which yields $\mathcal{R}_q=\mathcal{R}_\sigma$ \cite{Hauptmann2017}.  If $\sigma$ is a nonunitary constant, the problem can be scaled by that constant.  If the conductivity is not constant near $\bndry$, a transition zone can be added, see \cite{Nachman1988} for further details.  Note that $\mathcal{R}_{(q=0)}=\mathcal{R}_{(\sigma=1)}$.  Updating the ND maps from $q$ to $\sigma$ in \eqref{eq:texp_ND_q} gives

\vspace{-1em}
{\footnotesize
\begin{equation}\label{eq:texp_ND_sig}
   \texpND(\xi,\zeta) =  \underset{\bndry}{\int} e^{-ix\cdot(\xi+\zeta)} (\xi+\zeta)\cdot\nu \left(\mathcal{R}_\sigma - \mathcal{R}_1\right) e^{ix\cdot\xi}(\zeta\cdot\nu)\;dS.
\end{equation}
}
As $\texpND(\xi,\zeta)$ is a Born approximation to $t(\xi,\zeta)\approx\hat{q}$, one can then proceed with the remainder of the algorithm (Steps~(iii) and (iv)) to recover the conductivity, bypassing Step~(i) completely as is common in the DN version of the $\texp$ approach.

The steps of the $\texpND$ algorithm for full boundary data are:
\[\left(\mathcal{R}_\sigma,\mathcal{R}_1\right)\overset{1}{\longrightarrow}\texpND(\xi,\zeta) \overset{2}{\longrightarrow}\qexp \overset{3}{\longrightarrow}\sigexp\]
 \begin{enumerate}
 \item Compute the approximate scattering data $\texpND$ for each $\xi\in\R^3$, where $|\xi|\leq T_\xi$, fix $\zeta\in\mathcal{V}_\xi$  using \eqref{eq:texp_ND_sig}.

 \item Recover the approximate potential $\qexp(x)$ for $x\in\Omega\subset\R^3$ using \eqref{eq:t-q} with $\texpND$.

 \item Recover the approximate conductivity $\sigexp$ by solving \eqref{eq:t-QtoSigma} with $\qexp$ and evaluating $\sigexp(x)=\left(\tilde{u}(x)\right)^2$.
 \end{enumerate}
The scattering data is computed on a finite ball of radius $T_\xi$ which acts as a low-pass filtering of the nonphysical nonlinear scattering data and helps to stabilize the method in the presence of noise.  Again, this formulation assumes that $\sigma\equiv 1$ in a neighborhood of $\bndry$.  If instead $\sigma\equiv \sigma_b$ constant, following \cite{Isaacson2004}, the problem can be scaled by instead using $\sigma_b\mathcal{R}_\sigma$ and re-scaling at the end: $\sigma(x)=\sigma_b \left(\tilde{u}(x)\right)^2$.  In practice, the true value for $\sigma_b$ is often unavailable and replaced by the best constant conductivity fit to the measured data, $\sigBest$, given by  
\begin{equation}\label{eq-gam-best}
\sigma_{\mbox{\tiny{best}}}=\frac{\sum_{k=1}^K\sum_{\ell=1}^LU_\ell^kU_\ell^k}{\sum_{k=1}^K\sum_{\ell=1}^LU_\ell^kV_\ell^k},
\end{equation}
where $U_\ell^k$ is the $k^{th}$ simulated voltage pattern measured on electrode $\ell$ with a homogeneous conductivity of $1$ S/m and $V_\ell^k$ is the $k^{th}$ voltage pattern measured on electrode $\ell$ for the conductivity $\sigma(x)$, as described in \cite{Isaacson2004}.

\normalsize
\subsubsection{Partial boundary data}\label{ssec:pbdry_texp}
When electrodes are placed only on a subset $\Gamma\subseteq \partial\Omega$, only current and voltage measurements are available on $\Gamma$ and hence only an ND map $\mathcal{R}_\sigma^\Gamma$  is available rather than the ND map for all of $\bndry$.  In this work, we proceed by replacing the boundary integral over $\bndry$ in \eqref{eq:texp_ND_sig} with an approximation, computed only over the subset $\Gamma$,

\vspace{-1em}
{\footnotesize
\begin{equation}\label{eq:texp_ND_pd}
   \texpND_{,\Gamma}(\xi,\zeta) =  \int_\Gamma e^{-ix\cdot(\xi+\zeta)} (\xi+\zeta)\cdot\nu \left(\mathcal{R}_\sigma^\Gamma - \mathcal{R}_1^\Gamma\right) e^{ix\cdot\xi}(\zeta\cdot\nu)\;dS.
\end{equation}
}
Clearly this is a rough approximation to $\texpND$ as information on $\bndry\setminus\Gamma$ is absent and may contain important details.  Alternative methods, such as `extended electrodes' \cite{Hyvonen2009,Hauptmann2017a} are the subject of future work, and the 2D approach of \cite{Hauptmann2017}, which accounts for the missing information by using a differencing approach, does not naturally extend to the 3D setting as the 3D current patterns are significantly more complicated for the varying electrode configurations.  While spherical harmonics offer a possible path forward, this comes at the cost of electrode configuration flexibility.  An in-depth analysis of the error introduced by this simplified approximation is the subject of future work. For this proof-of-concept study, this approximation performed adequately and these methods have been demonstrated, on full boundary data \cite{Hamilton2021}, to be remarkably robust to incorrect domain modeling, and partial data here.

\subsubsection{Difference Imaging}\label{ssec:diff_im_texp}
For difference imaging, the map $\mathcal{R}_1$ is replaced with $\mathcal{R}_{\mbox{\tiny ref}}$, corresponding to a reference set of voltage measurements $V_{\mbox{\tiny ref}}$ and 

\vspace{-1em}
{\footnotesize
\begin{equation}\label{eq:texp_ND_diff}
   \texpNDdiff(\xi,\zeta) =  \int_{\bndry} e^{-ix\cdot(\xi+\zeta)} (\xi+\zeta)\cdot\nu \left(\mathcal{R}_\sigma - \mathcal{R}_{\mbox{\tiny ref}}\right) e^{ix\cdot\xi}(\zeta\cdot\nu)\;dS.
\end{equation}
}
Again, if $\sigma\equiv \sigma_c\neq 1$ near $\bndry$, both maps $\mathcal{R}_\sigma$ and $\mathcal{R}_{\mbox{\tiny ref}}$ are then scaled by the $\sigma_c$, or $\sigBest$, as described above in Section~\ref{sec:texp}. Then, $\sigdiff(x) = \sigBest\left(\left(\tilde{u}(x)\right)^2 - 1\right)$.  When using partial boundary measurements, the formulation can be updated to take place over $\Gamma$ rather than $\bndry$ as above in \S~\ref{ssec:pbdry_texp}.
\subsection{Calder\'on's method}
Calder\'on's method, \cite{Calderon1980}, relies on assuming the conductivity, $\sigma(x)$, is a small perturbation, $\delta\sigma(x)$, from a constant background, $\sigma_1$, so that $\sigma(x)=\sigma_1+\delta\sigma(x)$, where $\sigma_1=1$ with absolute imaging. For more details, see \cite{Calderon1980, Bikowski2008,Muller2017,Hamilton2021,Hamilton2022}. Calder\'on's method recovers the conductivity perturbation by taking the inverse Fourier transform of a boundary integral containing current and voltage data. The conductivity can then be approximated by adding back the constant background, $\sigma_1$. 

Under the traditional DN-map formulation, \cite{Hamilton2021,Hamilton2022}, the three steps of Calder\'on's method are
\begin{enumerate}
\item Use $\Lambda_\sigma$ and $\Lambda_1$ to approximate 

\vspace{-1em}
{\footnotesize
\begin{equation}\label{eq:CaldFhat}
 \widehat{\delta\sigma}(z)\approx\Fhat(z):=-\frac{1}{2\pi^2|z|^2}\int_{\partial\Omega}u_1\left(\Lambda_\sigma-\Lambda_1\right) u_2dS(x),
\end{equation}
}
where where $u_1:=e^{i\pi(z-ia)\cdot x}$, $u_2:=e^{i\pi(z+ia)\cdot x}$, and $z$ and $a$ satisfy 
\(z,a\in\R^3, |z|=|a|,\text{ and }z\cdot a=0.\) 

\item Take the inverse Fourier transform of $\Fhat(z)$: 
{\scriptsize
\begin{equation}\label{eq:cal-FhattoGammaCart}
\delta\sigma^{\mbox{\tiny{\textbf{CAL}}}}(x) \approx\mathcal{F}^{-1}\{\Fhat(z)\}(x)=\int_{\mathbb{R}^3}{\Fhat(z)\hat{\eta}\left(\frac{z}{y}\right)e^{-2\pi i(x\cdot z)}dz},
\end{equation}}
where $\hat{\eta}\left(\frac{z}{y}\right)$, is a mollifier to reduce Gibbs phenomenon \cite{Calderon1980}. 

\item Add the background to the perturbation to recover the approximate conductivity, $\sigma^{\mbox{\tiny{\textbf{CAL}}}}(x)$:
\begin{equation}\label{eq:sigCal_pert}
\sigma^{\mbox{\tiny{\textbf{CAL}}}}(x)=\sigma_{1}+\delta\sigma^{\mbox{\tiny{\textbf{CAL}}}}(x).
\end{equation}
\end{enumerate}
Note that Calder\'on's method uses a different definition of the Fourier transform than the $\texp$ method, but each method's use is self-consistent. As with the $\texp$ method, a low-pass filter is applied to $\Fhat$ in \eqref{eq:cal-FhattoGammaCart}, by computing $\Fhat$ on a ball of radius $T_{z_1}$, and then searching within a spherical shell from $T_{z_1}$ to $T_{z_2}$ for similar magnitude values of $\Fhat$ as given by equations (2.10) and (2.11) of \cite{Hamilton2022}. The algorithm for absolute imaging is 
\[\left(\Lambda_\sigma,\Lambda_1\right)\overset{1}{\longrightarrow}\Fhat(z) \overset{2}{\longrightarrow}\delta\sigma^{\mbox{\tiny{\textbf{CAL}}}}\overset{3}{\longrightarrow}\sigma^{\mbox{\tiny{\textbf{CAL}}}}\]
\normalsize

As shown in \cite{Bikowski2011}, there is an alternative, equivalent, formulation of Calder\'on's reconstruction method using $\texp$. The DN-map formulation of $\texp$ is
\begin{equation}\label{eq:texp}
\texp(\xi,\zeta)=\int_{\bndry} e^{-ix\cdot(\xi+\zeta)}(\Lambda_\sigma-\Lambda_1)e^{ix\cdot\zeta}d S(x),
\end{equation}
where $\zeta\in V_\xi$ as defined in \S\ref{sec:texp}. Noting that the criteria for choices for $\zeta, \xi, z$ and $a$ are imposed to guarantee that the exponentials are harmonic, the connection between $\texp$ and $\Fhat$ lies in the relationship between these variables. As $z,a\in\R^3$, we will compute $\zeta$ and $\xi$ from them. Let
\begin{equation}\label{eq:xi_and_zeta_from_za}
\zeta=\pi(z+ia)\quad\text{and}\quad
\xi=-2\pi z.
\end{equation}
Substituting \eqref{eq:xi_and_zeta_from_za} into the exponentials in \eqref{eq:texp} yields
\begin{equation}\label{eq:texp_za}
\texp(z)=\int_{\bndry} e^{i\pi(z-ia)\cdot x}(\Lambda_\sigma-\Lambda_1)e^{i\pi(z+ia)\cdot x}d S(x).
\end{equation}
It is clear to see the relationship between \eqref{eq:CaldFhat} and \eqref{eq:texp_za} as
\begin{equation}\label{eq:Fhat_texp}
\Fhat(z)=-\frac{1}{2\pi^2|z|^2}\texp(z).
\end{equation}
Since the DN-map formulation of $\texp$ is equivalent to $\texpND$, it follows that the ND-formulation of $\Fhat$ is
{\begin{align}\label{eq:Fhat_ND}
\FhatND(z)&=-\frac{1}{2\pi^2|z|^2}\texpND(z)\nonumber\\
&=-\frac{1}{2\pi^2|z|^2}\int_{\bndry} (\partial_\nu u_1)(\mathcal{R}_1-\mathcal{R}_\sigma)\partial_\nu u_2d S(x)\nonumber\\
&=-\frac{1}{2|z|^2}\int_{\bndry} \tilde{u}_1(\mathcal{R}_\sigma-\mathcal{R}_1)\tilde{u}_2d S(x),
\end{align}}
where $\tilde{u}_1=((z-ia)\cdot \nu)u_1$ and $\tilde{u}_2=((z+ia)\cdot \nu)u_2$.

The algorithm for the $\FhatND$ Calder\'on's method is then:
\[\left(\mathcal{R}_\sigma,\mathcal{R}_1\right)\overset{1}{\longrightarrow}\FhatND(z) \overset{2}{\longrightarrow}\delta\sigma^{\mbox{\tiny{\textbf{CAL}}}}\overset{3}{\longrightarrow}\sigma^{\mbox{\tiny{\textbf{CAL}}}},\]
where the first step of Calder\'on's method uses \eqref{eq:Fhat_ND} instead of \eqref{eq:CaldFhat} and steps 2 and 3 remain the same.\\
\subsubsection{Partial boundary data}\label{ssec:pbdry_Cald}
Due to the connection between $\Fhat$ and $\texp$, Calder\'on's method likewise must rely on the ND map $\mathcal{R}_\sigma^\Gamma$ resulting in the use of 
\begin{equation}\label{eq:Fhat_ND_pd}
    \FhatND_{,\Gamma}(z)=-\frac{1}{2|z|^2}\int_{\Gamma} \tilde{u}_1(\mathcal{R}_\sigma^\Gamma-\mathcal{R}_1^\Gamma)\tilde{u}_2d S(x)
\end{equation}
in the first step of the method as an approximation to the boundary integral in $\FhatND$. As with the $\texp$ method, if the background conductivity is not 1, we similarly scale the ND maps and the final $ \sigma^{\mbox{\tiny{\textbf{CAL}}}}$ by $\sigma_{\mbox{\tiny{best}}}$ as computed by \eqref{eq-gam-best}
\subsubsection{Difference imaging}\label{ssec:diff_im_Cald} As with the $\texp$ method, the map $\mathcal{R}_1$ is replaced with $\mathcal{R}_{\mbox{\tiny ref}}$ yielding
\begin{equation}\label{eq:Fhat_ND_pd_diff}
    \FhatNDdiff(z)=-\frac{1}{2|z|^2}\int_{\bndry} \tilde{u}_1(\mathcal{R}_\sigma-\mathcal{R}_{\mbox{\tiny ref}})\tilde{u}_2d S(x).
\end{equation}
Using \eqref{eq:Fhat_ND_pd_diff} in step 1 of Calder\'on's method and stopping after step 2 results in the difference image $\delta\sigcal$. If the background conductivity is not 1, the same scaling by $\sigBest$ is used here.

\subsection{Comparison Methods}\label{sec:othermethods}

To assess the performance of the CGO-based methods, the absolute imaging reconstructions are compared to total variation regularized non-linear least squares solutions, and the difference imaging reconstructions are compared to traditional linear difference imaging reconstructions.

\subsubsection{Total variation regularized absolute imaging}\label{sec:TV}
A popular numerical approach for absolute imaging is total variation (TV) regularized non-linear least squares minimization
\begin{equation}
  \hat\sigma = {\rm arg} \min_{\sigma>0}\{\Vert L_e \left( V-U(\sigma,z_\ast) \right)\Vert^2
+ \alpha {\rm TV}_\beta (\sigma)\},
\label{equ:minTV}
\end{equation}
where $U(\sigma,z)$ is the finite dimensional forward map that gives electrode voltages for given domain conductivities $\sigma$ and electrode contact impedances $z$, $L_e$ is the Cholesky factor of the noise precision matrix $C_e^{-1}$ of $e$ so that $L_e^{\rm T}L_e = C_e^{-1}$, scalar valued $\alpha$ is the regularization weight parameter and ${\rm TV}_\beta (\sigma)$ is the (smoothened) TV regularization functional \cite{Rudin1992}
\begin{equation}\label{equ:TV}
{\rm TV}_\beta (\sigma) = \int_\Omega \sqrt{\| \nabla \sigma \|^2 + 
\beta }\ {\rm d} r,
\end{equation}
where the scalar valued $\beta$ is the (fixed) smoothing parameter. An initialization step of the minimization is used to give a spatially constant $\sigma_0$ and the contact impedances $z_\ast$ on the current injecting electrodes with 1+$L$ parameter non-linear least squares estimation
\begin{equation}
  (\sigma_0, z_\ast) = {\rm arg} \min_{\sigma_c, z >0} \{\Vert L_e \left( V-U(\sigma_c,z) \right)\Vert^2\},
\label{equ:initial_est}
\end{equation}
where the scalar $\sigma_c \in \mathbb{R}$ is the coefficient of a spatially constant conductivity image $\sigma_c \mathbf{1} \in \mathbb{R}^N$ and $z\in \mathbb{R}^L$ where $L$ is the number of electrodes. Both \eqref{equ:minTV} and \eqref{equ:initial_est} are solved using a lagged Gauss-Newton method equipped with a line search algorithm. The line search uses bounded minimization to enforce the non-negativity $\sigma > 0$. For more details of the method, see \cite{Toivanen21}.

The forward model $U(\sigma,z)$ in \eqref{equ:minTV} is based on a finite element (FEM) discretization of the complete electrode model \cite{Somersalo1992}. For details of the FEM implementation, see \cite{Vauhkonen1998a}. 

\subsubsection{Linear Difference Imaging}\label{sec:LD}
Linear difference imaging, see e.g. \cite{Adler2009greit}, reconstructs the change in conductivity between two measurements $(V_1; V_2)$. The reference linear difference imaging approach of this paper uses linearized approximations of the observation models
\begin{IEEEeqnarray}{rCl}
    V_1 &\approx& U(\sigma_0,z_\ast) + J_\sigma (\sigma_1-\sigma_0) + e_1 \label{equ:lin_model1}\\
    V_2 &\approx& U(\sigma_0,z_\ast) + J_\sigma (\sigma_2-\sigma_0) + e_2,
    \label{equ:lin_model2}
\end{IEEEeqnarray}
where the Jacobian matrix $J_\sigma$ of $U(\sigma,z_\ast)$ is evaluated at $\sigma = \sigma_0$, and the spatially constant $\sigma_0$ and the contact impedances $z_\ast$ on the current injecting electrodes are obtained from \eqref{equ:initial_est}. With the linearized models \eqref{equ:lin_model1} and \eqref{equ:lin_model2}, the difference in the measurements becomes
{\small
\begin{IEEEeqnarray}{rCl}
    \nonumber &\delta V &= V_2 - V_1\\
    \nonumber &&= \left(U(\sigma_0) + J_\sigma (\sigma_2-\sigma_0) + e_2\right) - \left(U(\sigma_0) + J_\sigma (\sigma_1-\sigma_0) + e_1\right) \\
    \label{equ:dV}
     &&= J_\sigma \delta\sigma + \delta e,
\end{IEEEeqnarray}
}
where $\delta \sigma = \sigma_2 - \sigma_1$ and $\delta e = e_2 - e_1$.
Using \eqref{equ:dV}, the change in conductivity $\delta \sigma$ can be reconstructed using the difference data $\delta V$, and a popular choice for obtaining $\delta\sigma$ is to use (generalized) Tikhonov regularization with a smoothness promoting regularization functional
\begin{equation}
\label{equ:minLD}
  \hat{\delta\sigma} = \arg \min_{\delta \sigma} \left\{ || L_{\delta e}(\delta V - J_\sigma \delta\sigma) ||^2 + ||L_p \delta \sigma||^2
  \right\},
\end{equation}
where $L_{\delta e}$ is the Cholesky factor of the noise precision matrix of $\delta e$ so that $L_{\delta e}^{\rm T}L_{\delta e} = C_{\delta e}^{-1} = \left(C_{e_1} + C_{e_2}\right)^{-1}$. In this paper, the regularization matrix $L_p$ is constructed utilizing a distance based covariance function \cite{lieberman2010parameter} and by setting $L_p^{\mathrm{T}}L_p = C_p^{-1}$, where the (prior covariance) matrix $C_p$ is constructed as
{\small
\begin{align}
\label{equ:GammaSmooth}
  C_{p}(i,j) = \mathrm{std(}\sigma)^2\exp\left(-\frac{\|x_i - x_j \|^2}{2a^2}\right), \  i, j = 1,\ldots, N,
\end{align}
}
where $\mathrm{std}(\sigma)$ controls the standard deviation and $a$ the correlation length of conductivity.


\section{Testing Data \& Evaluation Metrics}\label{sec:metricstesting}
How well a reconstruction method performs depends heavily on the task at hand.  Is target localization, volume, or conductivity value the most important?  Does your ranking change if the computational cost is too high?  Is robustness to noise or and/or domain modeling errors most important?  The easy answer is that we seek a method that excels in all categories.  In reality, one method rarely dominates all categories.  Here we highlight localization error and computational cost, while exploring test cases to push these partial boundary data CGO methods to their breaking points.
\subsection{A Rectangular Prism}\label{subsec:test1}
To illustrate the effect of missing electrode information on a portion of the boundary, data was simulated on a rectangular prism box. The box was simulated to be 17cm $\times$ 25.5cm $\times$ 17cm ($x_1\times x_2\times x_3$) with the origin in the center of each dimension.  Three coverage scenarios were considered:
\begin{itemize}
    \item Full boundary data: 32 electrodes were placed all around the box (Fig~\ref{subfig:32_elecs}).
    \item Partial boundary data 1: 28 electrodes were placed on the box, leaving one face without electrodes (Fig~\ref{subfig:28_elecs}).
    \item Partial boundary data 2: 20 electrodes were placed on the box, leaving one face without electrodes along with an additional ring of 8 electrodes removed from the 28 electrode case (Fig~\ref{subfig:20_elecs}).
\end{itemize}

\begin{figure}[h!]
    \centering
    \includegraphics[width=0.95\columnwidth]{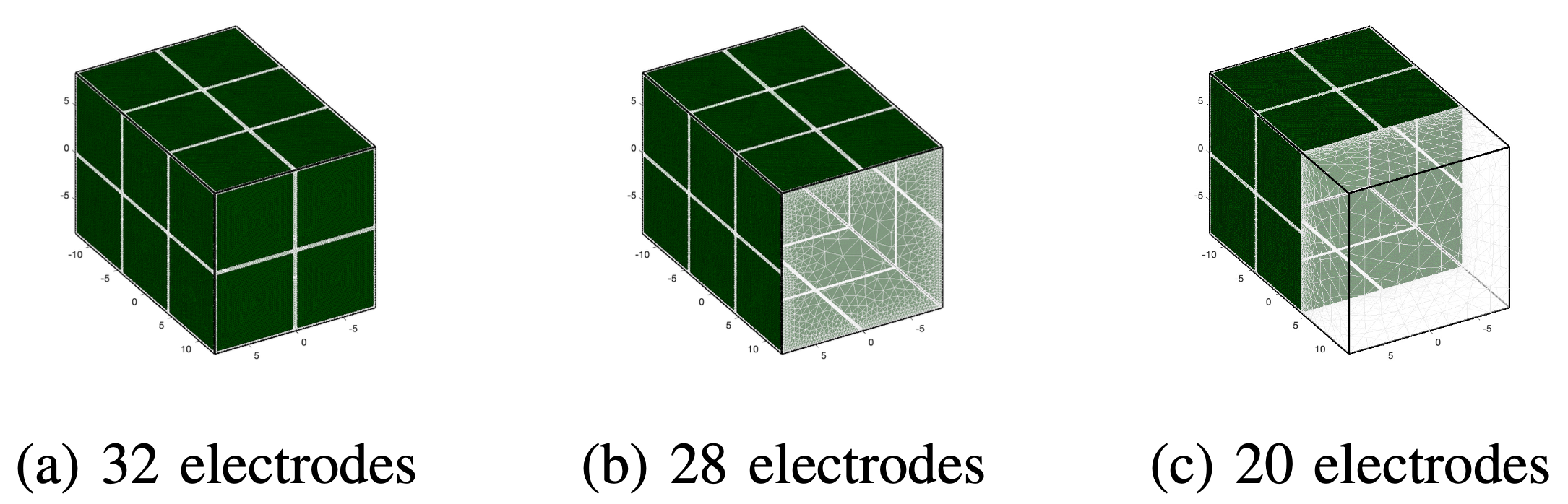}
    \caption{Electrode configurations}
\end{figure}
Simulations for all cases in the rectangular prism domain were performed using EIDORS~\cite{EIDORS}. For all three electrode configurations, a ball of radius 3cm was simulated to ``move" along the $x_2$ direction of the box, while maintaining the same $x_1$ and $x_3$ coordinates, from $(4.25,-9,-4.25)$ to $(4.25,9,-4.25)$. Note that this direction of movement is from a location that always had electrode coverage to a location that had electrodes removed in both partial boundary scenarios. The simulated background conductivity was 25~mS/m and the ball had conductivity of 125~mS/m. For each electrode configuration, optimal current patterns were applied as described in \cite{Hamilton2022} with the 32 electrode pattern available on GitHub\footnote{See {\scriptsize\url{https://github.com/sarahjhamilton/open3D_EIT_data}}.}.

To test robustness to noise, Gaussian relative noise was added to the simulated voltages 
\begin{equation}\label{eq:noisyVolts}
    V_{\mbox{\tiny noisy}}^j = V^j +  \eta\;  \texttt{mean}\left(|V^j|\right)N^j,
\end{equation}
where $V^j$ denotes the voltage corresponding to the $j^{\mbox{\tiny th}}$ applied current pattern, $N^j$ the noise vector unique for each current pattern, and $\eta$ denotes the noise level. Here we consider three levels of added noise: $0.01\%$, $0.1\%,$ and $1\%$.  These levels roughly correspond to the SNR capabilities of the ACT5 system \cite{ACT5} at 96dB and the KIT4 system at 65dB \cite{nissinen2011compensation}.  The unrealistic $1\%$ noise case was included to push the limits of the methods.  Note that no noise is added to the simulated background 1 case used for absolute imaging as that is simulated in experimental data settings as well. 

\subsection{A Computational Head Model}\label{subsec:test2}
Next, we explored how well the methods performed on a human head shaped domain and partial boundary setup, motivated by developing EIT-based imaging for detection of stroke, or expansion of hemorrhagic stroke. We considered a three layer phantom with the highly resistive skull, as well as a simplified (and unrealistic) case with no skull, and tested how well the $\texp$ method held up in both settings for increasing levels of noise for both absolute and difference imaging.  

The conductivity values for the head model are based on \cite{gabriel1996dielectric} and were 0.06~S/m for white matter, 0.02~S/m for the skull and 0.06 S/m for the skin layer. The conductivity of the simulated hemorrhagic stroke was set to 0.70~S/m for the 1.5~cm radius spherical inclusion that approximately matches the median volume (14~ml) of an intracerebral hemorrhage \cite{Robinson2022}. The first test case has a homogeneous head of only white matter and a simulated hemorrhage. The second test case includes the highly resistive skull. The difference imaging test cases additionally use reference datasets that do not contain a hemorrhage, simulating a situation of detecting the growth of an intracerebral hemorrhage or occurrence of secondary hemorrhage during treatment of ischemic stroke. The simulated measurements were obtained numerically using a finite element method (FEM) approximation of the complete electrode model (CEM) \cite{Cheng1989, Somersalo1992}. For details of the FEM implementation, see \cite{Vauhkonen1998a}. To solve the FEM problem and generate the simulated voltage data, the head model domain was divided into 417,205 tetrahedral elements and 77,840 nodes, 32 circular electrodes with a diameter of 1~cm and contact impedance of 0.003~$\Omega$m$^2$ were modeled on the scalp,  and were used for 31 pairwise adjacent current injections of 1~mA at 25 kHz frequency and the resulting electrode potentials were recorded on the same 32 electrodes, resulting in 992 simulated measurements in total. Reconstructions using the $\texp$ method were computed on a mesh with 57,526 nodes. 

\subsection{Experimental Tank Data Mimicking a Stroke}
The experimental tank data was measured using the KIT5 stroke EIT prototype measurement device \cite{Toivanen21}. The measurement setup used a 3D printed geometrically realistic adult head shaped tank, and a geometrically and electrically realistic 3D printed skull, both based on \cite{Avery2017}. The tank was filled with saline of conductivity 0.41~S/m and either a conductive or a resistive cylindrical target was suspended in the saline. The conductive cylinder had a conductivity of 4.73~S/m and the resistive cylinder was completely insulating. Both cylinders had height and diameter of 39~mm. Sixteen circular electrodes of diameter 1~cm were affixed to the scalp on the head-shaped tank, and 16 pairwise adjacent current injections of 1~mA at 25~kHz frequency were applied, and the resulting electrode voltages recorded on all 16 electrodes, resulting in 256 voltage measurements in total.  Unlike the comparison reconstruction methods, the CGO-based methods ($\texp$ and Calder\'on) only used the first 15 current patterns, due to the linear independence requirement in the numerical solution method and discrete ND formulation \cite{Hamilton2021}. For image reconstruction, the comparison methods used a mesh with 41,626 nodes for the electric potential and 8,884 nodes for the conductivity.

\subsection{Incorrect Domain Modeling}
Lastly we considered how well the reconstruction algorithms handled incorrect domain modeling.  In practice, a patient-specific computational mesh may not be available, especially if the patient has no previous CT/MR scans on file.  Instead, a patient may be matched to an approximate, simplified, model based on easier to collect measurements.  For the head imaging scenario, frontal-to-occiput length, temple-to-temple length, crown-to-cranial base length, or head circumference could be used.  Two natural medical applications of this methodology include brain imaging for stroke classification or monitoring, or breast imaging for identifying possible cancerous tumors.  We sought to determine if incorrect domain modeling would cause the CGO methods to fail, given that a common practice in imaging is to match a patient to a simplified domain of a similar shape. To explore this, an {\em approximate} head mesh was created in EIDORS by merging a semi-ellipsoid with a right elliptical cylinder. 

\begin{figure}[h!]
\centering
\includegraphics[width=\columnwidth]{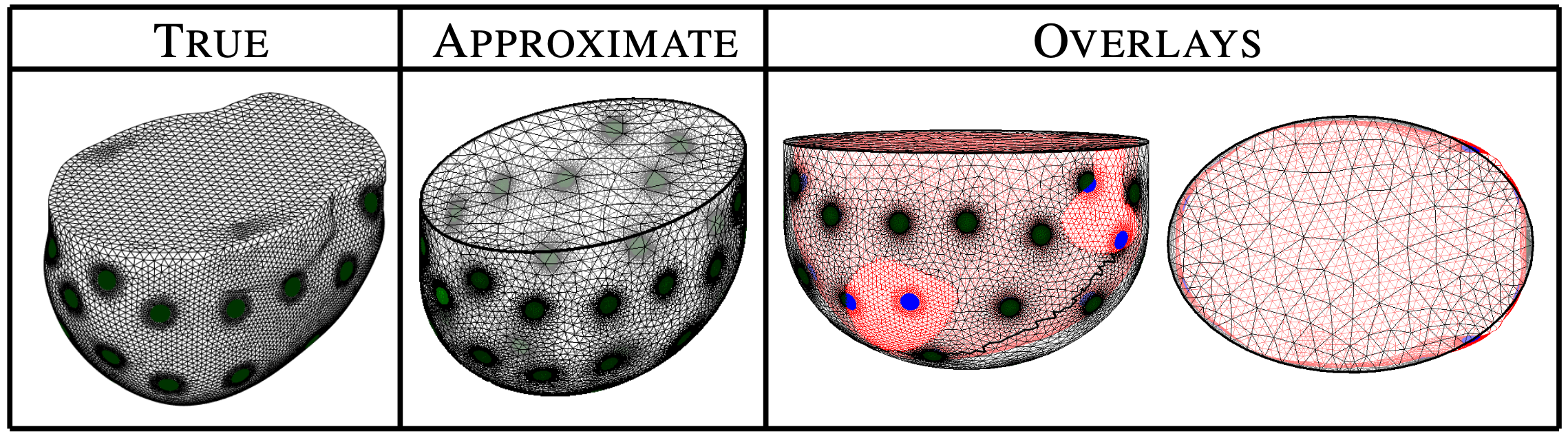}
%
%
%
%
%
%
%
%
%
    \caption{True versus approximate domain modeling. In the overlays, the true mesh is red with blue electrodes, and the approximate mesh is black with green electrodes.}
    \label{fig:domain_compare}
\end{figure}

The elliptical cylinder had major semi-axis length of 10~cm, minor semi-axis length of 7~cm, and its height ran from the origin to 4.5~cm. The semi-ellipsoid shared the major and minor semi-axes dimensions of the cylinder and had vertical/truncated semi-axis length of 8~cm, see Figure~\ref{fig:domain_compare}. This approximate head mesh was used to simulate data for a constant background of 1 S/m for absolute imaging. It was also used by all methods to reconstruct the conductivity when assuming the true domain was unknown. The correct centers and sizes of the electrodes were passed into the \texttt{place\_elec\_on\_surf} EIDORS function \cite{Grychtol2013}, which then re-meshed the domain with electrodes placed in approximate locations. The resulting mesh had 2,388 nodes and 11,083 elements. Approximate normals to the boundary at the electrodes were then calculated and used in the $\texp$ and Calder\'on's methods.

\begin{figure*}[t!]
\centering
\includegraphics[width=\textwidth]{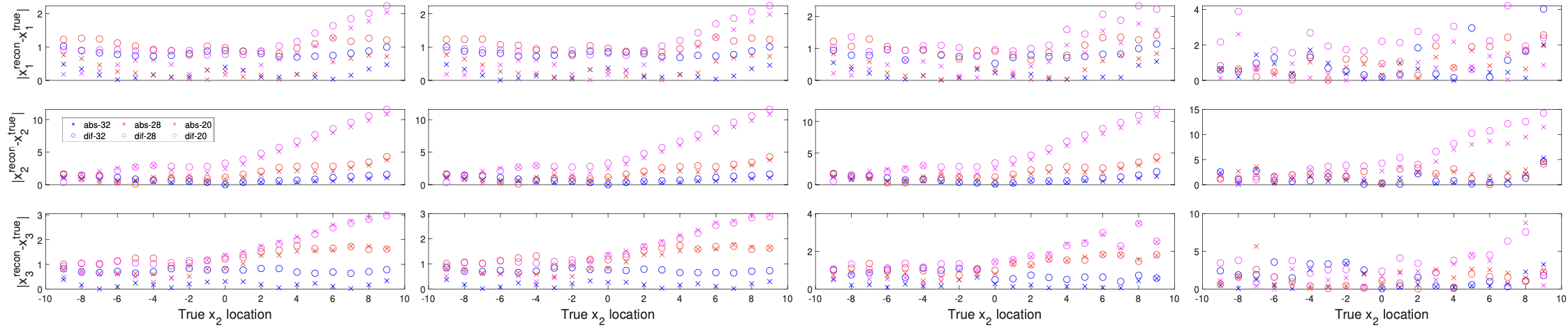}
%
%
     \caption{Difference image target location from truth (in cm) using Calder'on's method. Blue, red, and magenta represent results from the 32, 28, and 20 electrode configurations, while $\times$ and $\circ$ represent results from absolute and difference images.}\label{fig:targ_cents_NDonly_box}
\end{figure*}

\subsection{Evaluation Metrics}\label{ssec:metrics}
For the rectangular prism (\S\ref{subsec:test1}), we evaluate the effect of partial boundary data on localization of the reconstructed target. For this, we consider two metrics. First, as only the $x_2$ position of the target is changing, we calculate the absolute error in each coordinate as $|x_i^{\mbox{\tiny recon}}-x_i^{\mbox{\tiny truth}}|$. For both simulated data tests (\S\ref{subsec:test1} and \S\ref{subsec:test2}), we compute the localization error, LE, as in \cite{Hamilton2021}, to measure the distance between the centroids of the reconstructed targets and those of the true targets 
{\small
\begin{equation}\label{eq:LE}
        \text{LE}=\sqrt{(x_1^{\mbox{\tiny recon}}-x_1^{\mbox{\tiny truth}})^2+(x_2^{\mbox{\tiny recon}}-x_2^{\mbox{\tiny truth}})^2+(x_3^{\mbox{\tiny recon}}-x_3^{\mbox{\tiny truth}})^2}.
\end{equation}
}
An LE $=0$ would mean the center of the reconstructed target perfectly matches that of the true center.

For the experimental datasets, as the {\em true} location is unknown, we visually compare the reconstructions to the photographs taken during data collection. The image quality of the CGO methods is additionally assessed by direct comparison to the linear difference method (difference imaging) and TV method (absolute imaging), as well as their corresponding computational costs.  

\section{Results \& Discussion}\label{sec:results}
\subsection{The Effects of Partial Boundary Data vs. Noise}\label{sec:boxModel}
\begin{figure}
\centering
\includegraphics[width=\columnwidth]{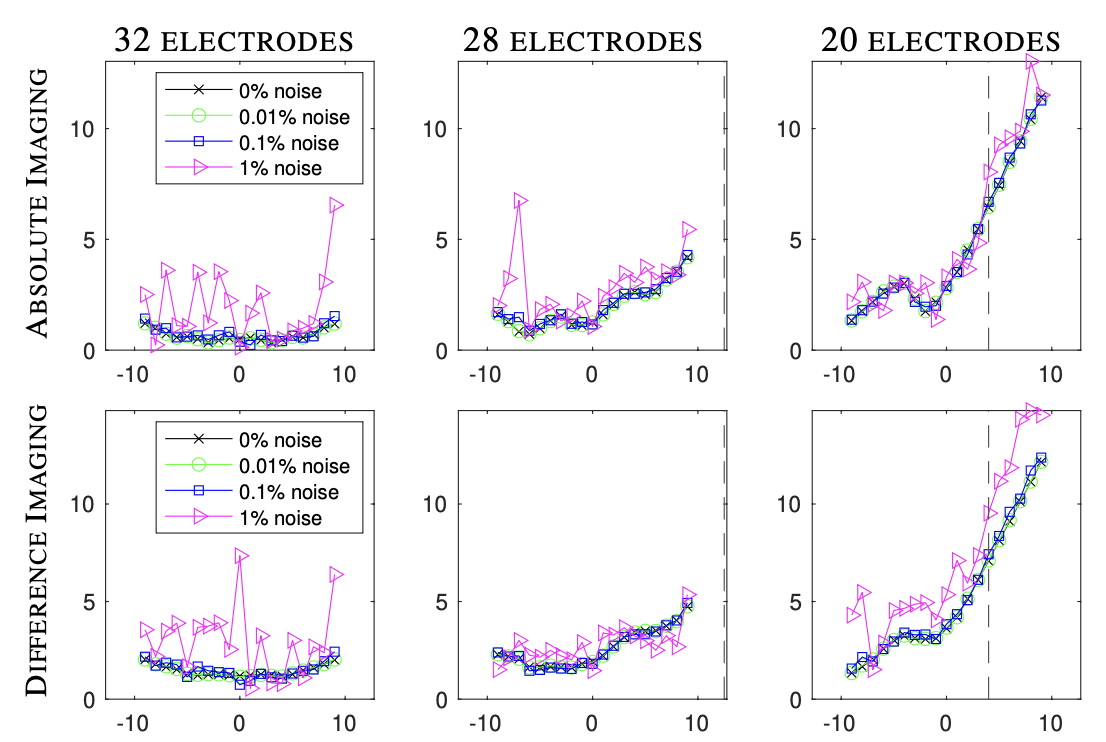}
    \caption{Localization error (LE) for Calder\'on's method reconstructions for all three electrode configurations. The vertical dashed line indicates the last $x_2$ location covered by an electrode. Black $\times$, green $\circ$, blue $\square$, and magenta $\triangleright$ indicate LE for $0\%, 0.01\%, 0.1\%,$ and $1\%$ noise, respectively.} 
    \label{fig:LE_box_ND}
\end{figure}

\begin{figure*}[h!]
\centering
\includegraphics[width=0.95\textwidth]{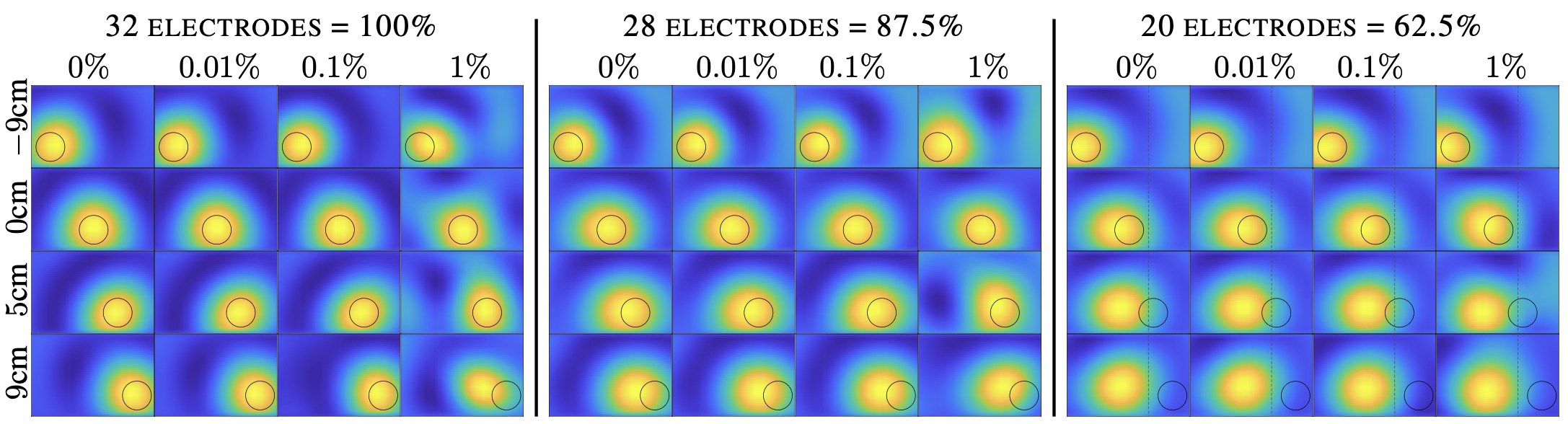}
\caption{Absolute images Calder\'on's method; cross-sections through the $x_3$ center of the true target. The black circle indicates the true target location. In cases with fewer electrodes, the right side is where electrodes are not present, as shown by the dashed vertical line.}
\label{fig:MovingBall_slices_All}
\end{figure*}

Focusing in on Calder\'on's method, using $T_{z_1}=0.8$ and $T_{z_2}=1.2$, Figure~\ref{fig:targ_cents_NDonly_box} displays the distance between the center of mass of the reconstructed target and the true center of the target for each of the coordinates for the 32, 28, and 20 electrode configurations for 0\%, 0.01\%, 0.1\%, and 1\% relative noise added to the voltages.  Results from both absolute and difference reconstructions are shown.  The localization error (LE) for each of the electrode configurations across the four noise levels for absolute (top) and difference (bottom) images is presented in Figure~\ref{fig:LE_box_ND}. The vertical dashed line in the 28 and 20 electrode configuration plots highlights the edge of electrode coverage, with no electrodes located to the right of that line.  Figure~\ref{fig:MovingBall_slices_All} shows absolute reconstructions, for all three electrode configurations across all four noise levels. The rows from top to bottom show the target moving from the end of the domain with full electrode coverage to the end that loses coverage in the 28 and 20 electrode configurations. A black circle is use to indicate the true location of the target. A vertical dashed line again highlights the end of the electrode coverage. 

Figures~\ref{fig:targ_cents_NDonly_box}-\ref{fig:MovingBall_slices_All} demonstrate that the target is locally reconstructed fairly well across noise levels until the $1\%$ noise case, which far exceeds the industry norm of $0.01\%$ or $0.1\%$. The larger source of error is seen in the configurations with missing electrodes. As the ball moved closer to the region without electrodes, the localization worsened and the target was pushed within the volume covered by electrodes.

%

\setlength{\unitlength}{1pt}

\subsection{Computational Head Model: Robustness to Noise}\label{sec:results_headmodel}
\begin{figure}[h!]    
\linethickness{.3mm}
    \centering

\includegraphics[width=\columnwidth]{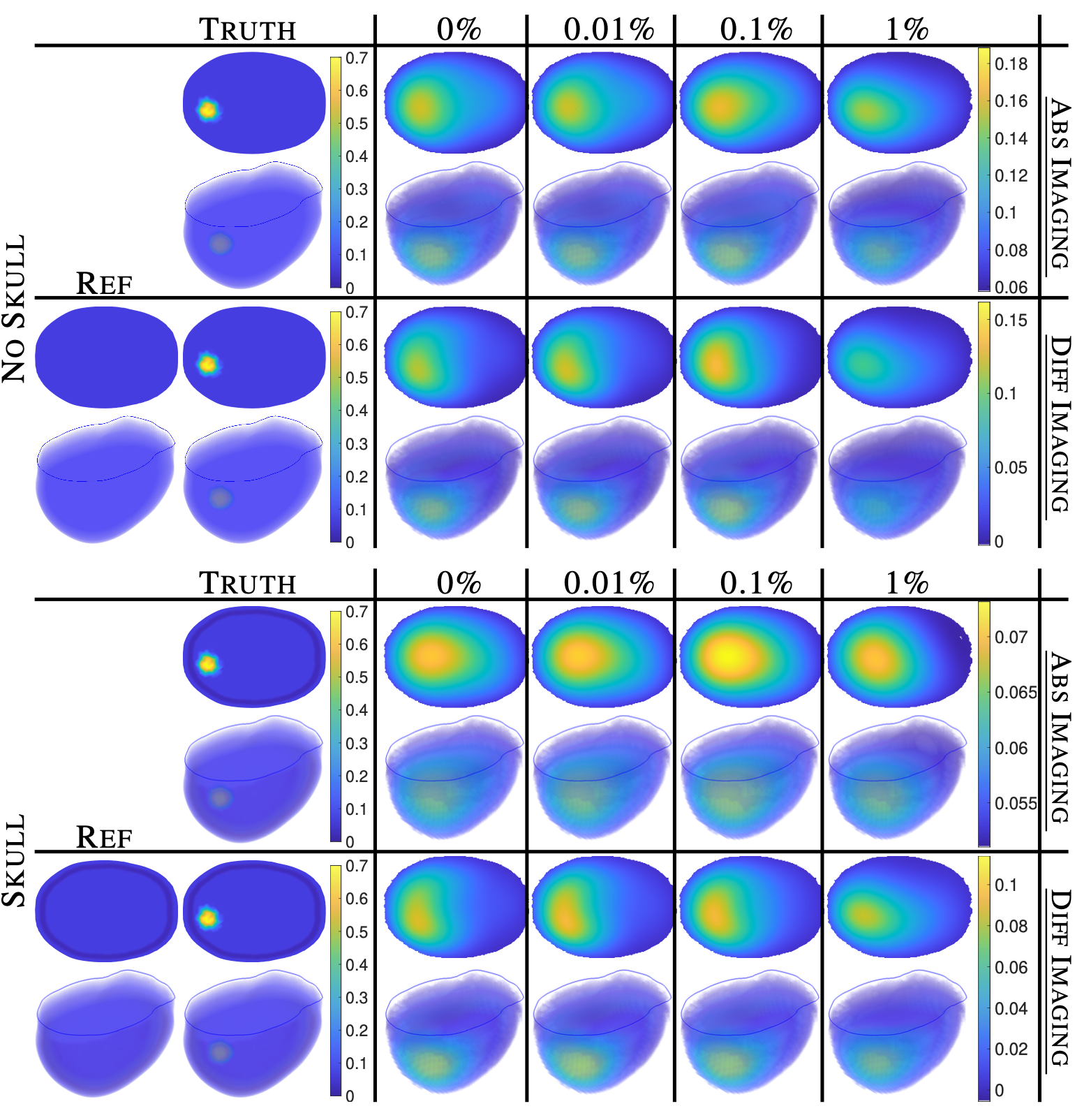}
    
    \caption{Comparison of the effect of noise on absolute imaging and difference imaging reconstructions using the $\texp$ method for computational head model with a hemorrhage {\bf with} and {\bf without} a skull.}
    \label{fig:HeadSims-noisy-ND-texp-NoMask-combined}    
\end{figure}
Figure~\ref{fig:HeadSims-noisy-ND-texp-NoMask-combined} shows the effect of noise in the voltage measurements for the absolute and difference imaging scenarios on the computational head model, with and without a skull.  Results are compared for 0\%, 0.01\%, 0.1\% and 1\% relative noise added to the non-reference voltage measurements.  Localization errors, along with approximate centroid coordinates, are reported in Table~\ref{table:LE_noisy}. Both the time difference and absolute images clearly identify the conductive target and hold up remarkably well to even high levels of noise.  As was observed in \cite{Hamilton2021} for the full boundary 3D CGO methods, the partial data $\texp$ reconstructions, for both absolute and time-difference imaging, contain the classic blurring inherent from the low-pass filtering employed for the computation of the scattering data \eqref{eq:texp_ND_sig}.  This blurring is significantly reduced if a thresholding post-processing mask is applied, which is common in many manuscripts but absent in the CGO literature (Fig.~\ref{fig:demo-masking}).  

From Table~\ref{table:LE_noisy} we see that the target is slightly better located for the cases without the skull, and the difference images had slightly better localization than the absolute images, as expected.  However, the results were remarkably stable for increasing levels of noise.  In all cases, the computed center of mass of the targets was lower in the $x_3$-direction than the truth.  As there is no voltage information coming from the flat portion (viewer's top) of the domain, this mislocalization is not unexpected and matches the behavior observed in \S~\ref{sec:boxModel} as the moving ball approached the portion of the domain with missing electrodes.  The recovered conductivity values underestimate the true contrast.  This is likely due to the low-pass filtering, which was stronger in the partial boundary data case than the full boundary cases studied in \cite{Hamilton2022}.  Here $T_\xi=12$ was used across noise levels for all cases except the absolute imaging with the skull, which used $T_\xi=5$.  Whether or not these CGO partial boundary methods would be capable of accurately recovering the contrast is the subject of future studies.  However, the standard methods for partial boundary EIT (e.g., linear difference and TV) are known to struggle to accurately recover the contrast, but are more time intensive. 

\begin{table}[h!]
\centering 
\setlength{\tabcolsep}{3pt}
\begin{tabular}{ll|rrrr|r|rrrr|}
\cline{3-6} \cline{8-11}
                                                          & \multicolumn{1}{c|}{}            & \multicolumn{4}{c|}{\sc Without Skull}                                                                            & \multicolumn{1}{l|}{\qquad} & \multicolumn{4}{c|}{\sc With Skull}                                                                                \\ \cline{2-11} 
\multicolumn{1}{l|}{}                                     & \multicolumn{1}{c|}{\sc Noise} & \multicolumn{1}{c|}{$x$}    & \multicolumn{1}{c|}{$y$}     & \multicolumn{1}{c|}{$z$}     & \multicolumn{1}{c|}{LE} & \multicolumn{1}{c|}{} & \multicolumn{1}{c|}{$x$}     & \multicolumn{1}{c|}{$y$}     & \multicolumn{1}{c|}{$z$}     & \multicolumn{1}{c|}{LE} \\ \hline
\multicolumn{1}{|l|}{\multirow{4}{*}{\rotatebox[origin=c]{90}{\sc Abs}}}   & 0\%                                & \multicolumn{1}{r|}{0.34} & \multicolumn{1}{r|}{-3.75} & \multicolumn{1}{r|}{-2.49} & 3.42                    &                       & \multicolumn{1}{r|}{-0.12} & \multicolumn{1}{r|}{-2.60} & \multicolumn{1}{r|}{-1.00} & 3.72                    \\ \cline{2-11} 
\multicolumn{1}{|l|}{}                                    & 0.01\%                           & \multicolumn{1}{r|}{0.28} & \multicolumn{1}{r|}{-3.70} & \multicolumn{1}{r|}{-2.48} & 3.45                    &                       & \multicolumn{1}{r|}{-0.12} & \multicolumn{1}{r|}{-2.61} & \multicolumn{1}{r|}{-1.00} & 3.70                    \\ \cline{2-11} 
\multicolumn{1}{|l|}{}                                    & 0.1\%                            & \multicolumn{1}{r|}{0.28} & \multicolumn{1}{r|}{-3.27} & \multicolumn{1}{r|}{-2.52} & 3.79                    &                       & \multicolumn{1}{r|}{-0.02} & \multicolumn{1}{r|}{-2.46} & \multicolumn{1}{r|}{-1.02} & 3.82                    \\ \cline{2-11} 
\multicolumn{1}{|l|}{}                                    & 1\%                             & \multicolumn{1}{r|}{0.94} & \multicolumn{1}{r|}{-3.30} & \multicolumn{1}{r|}{-2.07} & 3.41                    &                       & \multicolumn{1}{r|}{0.23}  & \multicolumn{1}{r|}{-2.69} & \multicolumn{1}{r|}{-1.11} & 3.57                    \\ \hline
\hline
\multicolumn{1}{|l|}{\multirow{4}{*}{\rotatebox[origin=c]{90}{\sc Diff}}} & 0\%                               & \multicolumn{1}{r|}{0.84} & \multicolumn{1}{r|}{-4.19} & \multicolumn{1}{r|}{-2.20} & 2.85                    &                       & \multicolumn{1}{r|}{0.69}  & \multicolumn{1}{r|}{-4.14} & \multicolumn{1}{r|}{-2.46} & 3.10                    \\ \cline{2-11} 
\multicolumn{1}{|l|}{}                                    & 0.01\%                           & \multicolumn{1}{r|}{0.86} & \multicolumn{1}{r|}{-4.18} & \multicolumn{1}{r|}{-2.18} & 2.84                    &                       & \multicolumn{1}{r|}{0.65}  & \multicolumn{1}{r|}{-4.14} & \multicolumn{1}{r|}{-2.44} & 3.09                    \\ \cline{2-11} 
\multicolumn{1}{|l|}{}                                    & 0.1\%                            & \multicolumn{1}{r|}{0.46} & \multicolumn{1}{r|}{-4.00} & \multicolumn{1}{r|}{-2.32} & 3.11                    &                       & \multicolumn{1}{r|}{0.56}  & \multicolumn{1}{r|}{-4.00} & \multicolumn{1}{r|}{-2.37} & 3.13                    \\ \cline{2-11} 
\multicolumn{1}{|l|}{}                                    & 1\%                             & \multicolumn{1}{r|}{0.82} & \multicolumn{1}{r|}{-3.79} & \multicolumn{1}{r|}{-2.08} & 3.04                    &                       & \multicolumn{1}{r|}{0.84}  & \multicolumn{1}{r|}{-3.59} & \multicolumn{1}{r|}{-2.28} & 3.33                    \\ \hline
\end{tabular}
\caption{Approximate target centroid locations (cm) and corresponding localization error across noise levels for absolute and time-difference imaging with and without a skull.  These metrics correspond to the data shown in Figure~\ref{fig:HeadSims-noisy-ND-texp-NoMask-combined}.  The true target was centered at $(1,-6,0)$.}
    \label{table:LE_noisy}
\end{table}

\subsection{Experimental Data: Correct Domain Modeling}\label{sec:results_tankData}
\begin{figure}[h!]
    \centering
\includegraphics[width=\columnwidth]{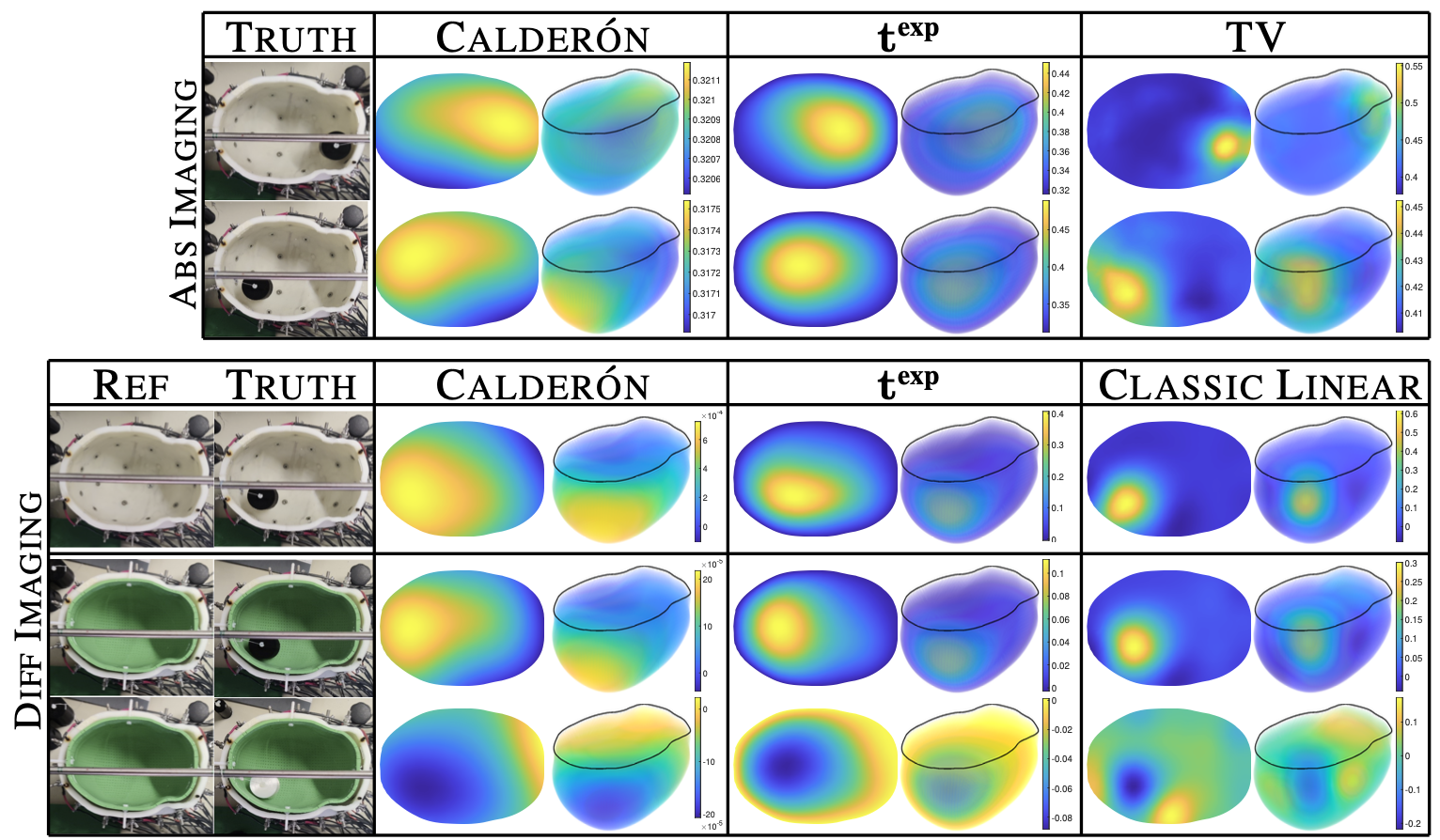}
    \caption{Absolute (top) and difference (bottom) image reconstructions (in S/m) from the CGO methods compared to the classic reconstruction methods.}
    \label{fig:UEF_June2023_correctDomain}
\end{figure}

 Datasets mimicking a hemorrhage (conductive black target in photograph) or ischemic stroke (resistive white target) are considered with and without the highly resistive (green) skull. Figure~\ref{fig:UEF_June2023_correctDomain} compares the results from the Calder\'on and $\texp$ methods, to the classic methods from \S~\ref{sec:othermethods} for absolute and time-difference imaging.  The reference voltages $V_{\mbox\tiny \text{ref}}$ for the difference images correspond to the no inclusion setups shown under the `Ref' label for each corresponding case.  The targets are well-located with and without the skull for the CGO difference imaging methods, with significantly more spreading in the Calder\'on reconstructions.  The classic linear difference and $\texp$ methods had the best contrast.  The results were similar for the absolute images.  Increasing the number of electrodes from 16 to 32 would likely improve the CGO reconstructions as was shown in \cite{Hamilton2021,Hamilton2022}. The regularization parameters used were: $T_\xi=5$ for the $\texp$ method for all cases except the no skull difference imaging example ($T_\xi=10$), $T_{z_1}\in[.3,.45]$ and $T_{z_2}\in[.65,.75]$ for Calder\'on's method, $\alpha = 10^{-4}$ and $\beta = 10^{-3}$ for TV, and $\mathrm{std}(\sigma) = 8.3$ mS/cm and $a = 1.3$ for the classic linear difference method.

\subsection{Experimental Data: Incorrect Domain Modeling}
\begin{figure}[h!]
\linethickness{.2mm}
    \centering

\includegraphics[width=\columnwidth]{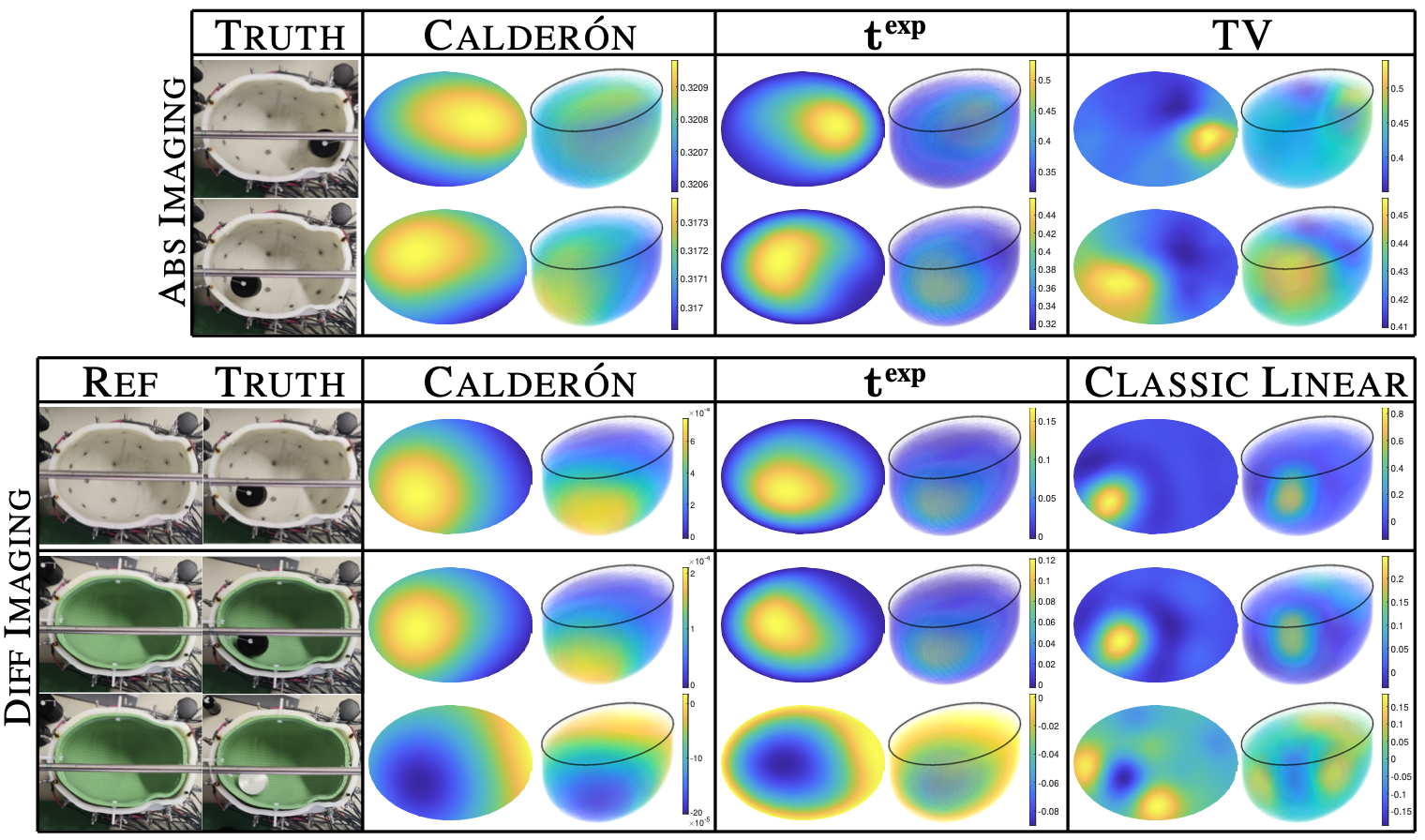}
    \caption{Absolute (top) and difference (bottom) image reconstructions from the CGO methods compared to the classic reconstruction methods with {\bf incorrect} domain modeling.}
    \label{fig:UEF_June2023-wrongDom}
\end{figure}

To further study the robustness of the CGO methods to incorrect domain modeling, we next used the approximate head model shown in Figure~\ref{fig:domain_compare}.  This incorrect modeling results in incorrect information for the computational domain used to recover $\sigma$ when solving the PDE in \eqref{eq:t-QtoSigma}, as well as computing $\sigBest$ in \eqref{eq-gam-best}.  Additionally, on this domain, the centers of the electrodes no longer match the experimental setup, which will lead to incorrect $x$ values in the computation of 
$\texpND$, $\texpNDdiff$, $\FhatND$, and $\FhatNDdiff$ and their associated normal vectors $\nu$ at those new centers. For absolute imaging, the $\texp$ and Calder\'on methods also require simulated voltages $V_1$ corresponding to the voltages that would occur on the same domain but with $\sigma=1$, which will not match the true domain.  The resulting reconstructions can be seen in Figure~\ref{fig:UEF_June2023-wrongDom}. Comparing this to Figure~\ref{fig:UEF_June2023_correctDomain} , we see that all methods considered held up quite well to this incorrect domain modeling a promising results for the use of a generic head model in stroke imaging.

\subsection{Further Discussion}\label{sec:discussion}
Given that the classic linear difference and total variation methods produced sharper localization of the targets, what benefit do these CGO-based methods provide?  A significant benefit comes in the form of computational cost.  

\subsubsection{Computational Cost}
None of the CGO reconstruction code in this project has been optimized for speed.  As such, Calder\'on's method produces absolute or time difference images in  approximately 0.2 to 0.5 seconds, when reconstructing on 2,021 voxels and 6-7 seconds for 23,319 nodes. Due to the poor spatial resolution of EIT, and inherent blur and Gibb's phenomenon in Calder\'on's method, there is little payoff in increasing the reconstruction mesh density.  The $\texp$ method took approximately 2 seconds to recover the conductivity on the 20,000+ mesh nodes used for this work.  The above timings correspond to code running in Matlab on personal laptops.  Note that for the CGO-based methods, there is no difference in time for absolute vs difference imaging.  

By comparison, the linear difference reconstructions for Figure~\ref{fig:UEF_June2023_correctDomain} with correct domain modeling required approximately 30 seconds each, on CSC Finland’s ePouta cloud server partition.  
This timing does not include the 12.33 second computational cost of computing the mesh and correlation distance-specific regularization matrix required for the correlation-based smoothness regularization as it can be computed once when the mesh is created. The computational cost of the TV (absolute) images, with correct domain modeling, ranged from 6 to 12 minutes. When using the the approximate head domain for incorrect modeling (Fig.~\ref{fig:UEF_June2023-wrongDom}), the linear difference reconstructions averaged 28 seconds each and the TV ranged from about 3 to 8 minutes.

Thus, the CGO-based methods produced absolute images 90-360 times faster than the TV approach, and about 15x faster than the comparison linear difference method, assuming the 2-second $\texp$ timing.

\subsubsection{Avenues for improving the CGO reconstructions}
\begin{figure}[h!]
\linethickness{.3mm}
    \centering
    \includegraphics[width=\columnwidth]{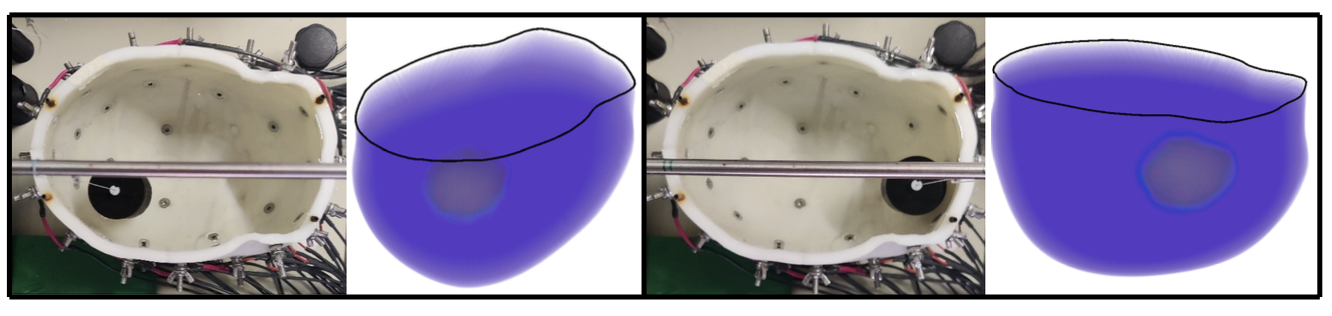}
%
%
%
%
%
%
		 \caption{Demonstration of thresholding post-processing of $\texp$ absolute images from Fig.~\ref{fig:UEF_June2023_correctDomain}.}
    \label{fig:demo-masking}
\end{figure}

Given that there are no sparsity priors enforced, intensity thresholding or post-processing applied to the CGO reconstructions, the blurring inherent in the CGO approaches could likely be mitigated.  Figure~\ref{fig:demo-masking} demonstrates one such approach where only conductivity values with large enough contrast to the background are retained. Pairing the CGO methods with a post-processing learning approach has been highly effective in 2D \cite{Hamilton2018_DeepDbar} \cite{hamilton2019beltrami}, and may improve the contrast in scenarios where a higher number of electrodes is not possible for the 3D partial data scenario. 

\section{Conclusions}\label{sec:conclusion}
In this work we developed, and numerically implemented, the first 3D partial boundary data CGO-based reconstruction algorithms for EIT.  We explored the limits of these new methods for large areas of missing electrode coverage, high levels of noise in the voltage data, anatomically realistic head models, experimental tank data, and incorrect domain modeling.  The CGO methods were able to reliably identify the targets, in the region of electrode coverage, for all cases considered.  The results for experimental data were compared to classic EIT reconstruction methods.  While the classic methods obtained better localization and sharpness of the imaged targets, the new CGO-based methods provided reconstructions in a small fraction of the time: 90-360x faster for absolute imaging and 15x faster for difference imaging.  The methods are exceptionally fast for 3D reconstruction and are likely capable of real-time imaging if optimized.  Combining the proposed approach, for example, with a learning-based post-processing holds potential for online applications, such as follow-up of hemorrhagic stroke or lung monitoring, where fast image reconstruction is needed for control or to guide treatment decisions.

\section*{Acknowledgment}\label{sec:acknowledgments}
The authors wish to acknowledge CSC - IT Center for Science, Finland, for computational resources.



\bibliographystyle{IEEEtran}
\small
\bibliography{bibrefs.bib}

\end{document}